\documentclass[a4paper,aps,nofootinbib,twocolumn]{revtex4}

\usepackage[english]{babel}
\usepackage{amssymb,amsthm}
\usepackage{amsmath}
\usepackage{amsfonts}
\usepackage{amssymb}
\usepackage{graphicx}
\allowdisplaybreaks[4]
\usepackage{natbib}
\begin{document}


\newcommand{\abs}[1]{\ensuremath{\left\lvert #1\right\rvert}}
\newcommand{\erf}[2][]{\ensuremath{\text{erf#1}\left(#2\right)}}
\newcommand{\eps}{\ensuremath{\varepsilon}}

\newcommand{\be}{\begin{equation}}
\newcommand{\ee}{\end{equation}}
\newcommand{\bs}{\begin{subequations}}
\newcommand{\es}{\end{subequations}}
\def\ee{\end{equation}}
\def\be{\begin{equation}}
\def\bdm{\begin{displaymath}}
\def\edm{\end{displaymath}}
\def\l{\left}
\def\r{\right}
\def\a{\alpha }

\newcommand{\se}{\ensuremath{_\perp}}

\newcommand{\Th}{\ensuremath{\varTheta}}
\newcommand{\Ga}{\ensuremath{\varGamma}}
\newcommand{\La}{\ensuremath{\varLambda}}

\newcommand{\uint}{\ensuremath{\int_{-\infty}^\infty}}

\newcommand{\f}[1]{\ensuremath{\boldsymbol{#1}}}
\newcommand{\im}{\ensuremath{\mathfrak{I\!m}}}

\newcommand{\pd}[2][]{\ensuremath{\frac{\partial #1}{\partial #2}}}
\newcommand{\df}{\ensuremath{\mathrm{d}}}

\newcommand{\etal}{ \emph{et\,al.}}

\title{Quantum plasma modification of the Lane-Emden equation for stellar structure}
\author{R. Schlickeiser}
\email{rsch@tp4.rub.de}
\author{I. Lerche}
\email{lercheian@yahoo.com}
\author{C. R\"oken}
\email{cr@tp4.rub.de}
\affiliation{Institut f\"ur Theoretische Physik, Lehrstuhl IV:
Weltraum- und Astrophysik, Ruhr-Universit\"at Bochum,
D-44780 Bochum, Germany}
\date{June 18, 2008}

\begin{abstract}
The proper quantum plasma treatment of the electron gas in degenerate stars such as white 
dwarfs provides an additional quantum contribution to the electron pressure. The additional pressure term 
modifies the equation for hydrostatic equilibrium, resulting in the quantum modified Lane-Emden equation for
polytropic equation of states. The additional pressure term also modifies the expression for the limiting
Chandrasekhar mass of white dwarfs. An approximate solution is derived of the quantum 
modified Lane-Emden equation for general polytropic indices, and it is demonstrated that the quantum corrections 
reduce the standard Chandrasekhar mass and enhance the white dwarf radius by negligibly small values only.
\end{abstract}

\maketitle

\section{Introduction}

From a plasma point of view the electron gas of density $n$ in degenerate stars such as white dwarfs 
is a nearly collisionless quantum plasma \cite{mo07} because its central temperature $T$ is lower than the 
Fermi temperature $T_F$, and the quantum coupling parameter $g_Q=E_c/E_F$ of the ratio of the 
Coulomb interaction energy $E_c\simeq e^2n^{1/3}$ to the Fermi energy $E_F$ is much smaller than unity. 
The white dwarf Sirius B has an average mass density of $3\cdot 10^6$ g cm$^{-3}$ and a central temperature of
$T_c=7.6\cdot 10^7$K yielding for the temperature ratio 

\be
\chi \equiv \frac{T_F}{T} = \l( 3 \pi^2 \r)^{2/3} \frac{\hbar^2n^{2/3}}{2 m k_BT}       
     = 3.65 \l( n \lambda _B^3\r)^{2/3}=55.7
\label{a1}
\ee
The ratio (\ref{a1}) indicates that the thermal de Broglie wavelength of individual plasma particles

\be
\lambda _B = \frac{\hbar}{mv_T}\simeq 3.9 n^{-1/3} \; , 
\label{a2}
\ee
is of the same order of magnitude as the mean distance between electrons $n^{-1/3}$. 
The thermal de Broglie wavelength roughly represents the spatial extension of a particle's wave function due 
to the quantum uncertainty. For $\lambda _B$ comparable to the interparticle distance, because of overlapping 
electron wave function extensions, individual electrons cannot be treated as pointlike 
particles as in the classical plasma description, and quantum effects become important. 
The smallness of the quantum coupling parameter $g_Q$ assures that collective mean field-effects dominate over 
binary collisions because the typical electron Fermi energy is much larger than the Coulomb interaction energy 
with neighbouring electrons.

Here we demonstrate that the proper quantum treatment of the electron gas based on the Wigner 
\cite{w32} distribution function changes the hydrostatic equilibrium equation in white dwarfs leading, however, to a 
negligible modification of the maximum Chandrasekhar mass of such systems. In the hydrodynamical equations the 
quantum effect \cite{mh01} yields an additional pressure term $P_Q$ to the 
classical pressure $P_c$, 

\be
P=P_c+P_Q=P_c+{\hbar ^2\over 2m}\l[(\nabla n^{1/2})^2-n^{1/2}\nabla ^2n^{1/2}\r]
\label{a3}
\ee
in terms of the electron density $n$. The additional pressure term $P_Q$ is not listed 
in Salpeter's \cite{s61} account of the pressure contributions to a zero-temperature degenerate Fermi gas of
non-interacting electrons. Here the modification to the standard Lane-Emden equation 
for polytropic gases and to the limiting Chandrasekhar mass are calculated.
\section{Quantum modified equations for hydrostatic stellar equilibria}
For a plasma at rest ($\vec{u}=0$) the hydrodynamical equation (\ref{rb10}) becomes 

\be
0=\vec{g}(\vec{x})+{1\over \rho }\nabla P_c
-{\hbar ^2Z\over 2mm_i}\nabla \l({\nabla ^2\rho ^{1/2}\over \rho ^{1/2}}\r)
\label{c1}
\ee
with the mass density $\rho =nm_i/Z$ where $m_i$ denotes the ion mass. 
Choosing spherical coordinates we obtain the quantum modified equation for hydrostatic equilibria
in stars as

\be
0=g(r)+{1\over \rho (r)}{dP_c\over dr}-
{\hbar ^2Z\over 2mm_i}{d\over dr}\left({\nabla _r^2\rho ^{1/2}(r)\over \rho ^{1/2}(r)}\right)
\label{c2}
\ee
with the operator 

\be
\nabla _r^2=r^{-2}{d\over dr}(r^2{d\over dr})
\label{c3},
\ee
and the usual gravitational acceleration 

\be
g(r)=G\mu (r)/r^2
\label{c4}
\ee
due to the enclosed mass 

\be
{d\mu \over dr}=4\pi r^2\rho (r)
\label{c5}
\ee
The mass inside the radius $r$ is given by 

\be
M(r)=\int_0^rdr^{'}\; {d\mu \over dr^{'}}=4\pi \int_0^rdr^{'}\; (r^{'})^2\rho (r^{'})
\label{c6}
\ee
so that the total mass is 

\be
{\cal M}=4\pi \int_0^Rdr\; r^2\rho (r)
\label{c6a}
\ee
Inserting Eq. (\ref{c4}) into Eq. (\ref{c2}), and differentiating with respect to $r$ immediately yields the 
quantum modified stellar structure equation 

\bdm
{d\over dr}\left[r^2\left({1\over \rho (r)}{dP_c\over dr}-
{\hbar ^2Z\over 2mm_i}{d\over dr}\left({\nabla _r^2\rho ^{1/2}(r)\over \rho ^{1/2}(r)}\right)\r)\r]
\edm
\be
=-4\pi Gr^2\rho (r)
\label{c7}
\ee
\subsection{Quantum modified Lane-Emden equation}
Following standard procedures \cite{c58,c83} we 
adopt the polytropic equation of state

\be
P_c=K\rho ^{(1+\alpha )/\alpha }
\label{c8}
\ee
and substitute $\rho =\rho _c y_{\alpha }^{\a}$ (with central density $\rho _c$) and $r=Ax$ with the constant 

\be
A=\l[{(\a +1)K\rho _c^{(1-\a )/\a}\over 4\pi G}\r]^{1/2}
\label{c11}
\ee
Eq. (\ref{c7}) then becomes the quantum modified Lane-Emden (QMLE) equation for a polytrope of index $\a$ 

\bdm
{1\over x^2}{d\over dx}\l(x^2{dy_{\a}\over dx}\r)+\, y_{\a}^{\a}=
\edm
\be
{\eta _{\a}\over x^2}{d\over dx}\l(x^2{d\over dx}{x^{-2}{d\over dx}\l(x^2{dy_{\a}^{\a/2}\over dx}\r)\over y_{\a}^{\a/2}}\r)
\label{c12}
\ee
where the dimensionless parameter 

\be
\eta _{\a}={\hbar ^2Z\over 8\pi Gmm_i\rho _cA^4}={2\pi G\hbar ^2Z\rho _c^{(\a-2)/\a}\over mm_iK^2(\a+1)^2}
\label{c13}
\ee
characterizes the strength of the quantum plasma modifications. The term on the right hand side of the 
QMLE equation (\ref{c12}) reflects the new quantum contribution from the 
Bohm potential. For $\eta _{\a}=0$ the QMLE-equation reduces to the 
standard Lane-Emden equation of stellar structure.

The QMLE equation (\ref{c12}) has been derived for flat Newtonian space-time ignoring general relativistic 
effects. For completeness in Appendix B we derive the relativistic generalization of the QMLE-equation.

\subsection{Quantum modified Chandrasekhar mass}
In terms of the substituted variables $x$ and $y_{\a}(x)$ the mass (\ref{c6}) inside the normalised radius $x$ reads 

\bdm
M(x)=4\pi \int_0^{Ax}dr^{'}\; (r^{'})^2\rho (r^{'})=
\edm
\be
4\pi \rho _c A^3\int_0^xd\xi \, \xi ^2y_{\a}^{\a}(\xi )
\label{c14}
\ee
Inserting the QMLE equation (\ref{c12}) yields 

\bdm
M(x)=-M_0\Bigl[x^2{dy_{\a}\over dx}-
\edm
\be
\eta _{\a}\l(x^2{d\over dx}{x^{-2}{d\over dx}\l(x^2{dy_{\a}^{\a/2}\over dx}\r)\over y_{\a}^{\a/2}}\r)\Bigr]
\label{c15}
\ee
with 

\be
M_0=4\pi \rho _cA^3
\label{c15a}
\ee
The first zero $x_1$ of the solution $y_{\a}(x_1)=0$ of the QMLE equation defines the size of the star.
In terms of $M_0$, with $Z=2$ and $m_i=Z\cdot 1836m$, the parameter $\eta _{\a}$ can be expressed in cgs-units as 

\be
\eta _{\a }={\hbar ^2Z\over 2Gmm_iAM_0}={5.4\cdot 10^3\over AM_0}=1.3\cdot 10^4{\rho _c^{1/3}\over M_0^{4/3}}
\label{c15b}
\ee
Scaling $M_0=3\chi M_{\rm sun}=6\chi \cdot 10^{33}$ g in solar masses and $A=a_9\cdot 10^9$ cm 
with the typical white dwarf mass and radius, we obtain very small values of the parameter 

\be
\eta _{\a}=9.2\cdot 10^{-39}(a_9\chi )^{-1} 
\label{c15c}
\ee
The small value of the parameter $\eta _{\a}$ reflects the the small ratio of the Fermi pressure and Bohm pressure terms 
(second and third term in Eq. (\ref{c2})). By replacing $(d/dr)$ by $1/L$, where $L$ denotes a typical length such as the
radius of the star, and using the relativistic expression for the Fermi pressure 
$P_c=(3\pi ^2)^{1/3}\hbar cn^{4/3}/4$, we find for the ratio of the two terms in terms of the 
Compton electron wavelength $\lambda _c=\hbar /(mc)=3.86\cdot 10^{-11}$ cm 

\be
{T_3\over T_2}=
{\hbar ^2n\over 2mP_cL^2}={2\over (3\pi ^2)^{1/3}}{\lambda _c\over L^2n^{1/3}}
\label{c15d}
\ee
For a white dwarf, typically we have $n\simeq 10^{30}$ cm$^{-3}$, so that with $L\simeq 10^9$ cm 
we estimate $T_3/T_2\simeq 2.5\cdot 10^{-39}$, in agreement with the estimate (\ref{c15c}).

While these rough order of magnitude estimates indicate that the Bohm pressure
term is likely a small contributor to modifications of the Chandrasekhar white dwarf mass limit,
nevertheless it is appropriate to figure out the effect quantitatively for two main reasons. First, it
can happen that the rough estimates given above do not uncover the total effect when it is worked out
quantitatively and so it could happen that a small correction has a profound effect on the solution to
the non-linear equation. Such effects are not unusual in physics and one truly needs to work out the details
in order to be sure that one has not overlooked some subtle component. Second, even when detailed
calculations show that the Bohm pressure effect is small, in accord with the rough estimates made, it is
then satisfying to note the confirmation of the rough estimates by detailed evaluation. For these two
reasons alone it is more than appropriate to evaluate the Bohm pressure effect although, as we will show,
the contribution is indeed small. The procedure for evaluating the contribution to the white dwarf mass
may also be of relevance to other astrophysical problems  and so, as a basic technique, it is also
appropriate to spell out the details of how one addresses such non-linear corrections.

Since the pioneering work of Chandrasekhar \cite{c31a,c31b,c31c} summarized in \cite{c58} it is well 
known that the interior of white dwarfs is supported
by the pressure of completely degenerate electrons. With increasing density the pressure becomes 
so high that the degeneracy becomes relativistic so that $P_c=1.244\cdot 10^{15}(\rho /\mu _e)^{4/3}$ which
corresponds to a polytrope index of $\a=3$. In this case Eq. (\ref{c15}) yields the 
quantum-modified Chandrasekhar mass limit 

\bdm
{\cal M}=M(x_1,\a =3)=
-{4\over \pi ^{1/2}}\l({K\over G}\r)^{3/2}
\edm
\be
\l[x^2{dy_3\over dx}-\, 
\eta _3x^2\l({d\over dx}{x^{-2}{d\over dx}\l(x^2{dy_3^{3/2}\over dx}\r)\over y_3^{3/2}}\r)\r]_{x_1}
\label{c16}
\ee
with 

\be
\eta _3=4.28\cdot 10^3{G^2\rho _c^{1/3}\over K^2}
\label{c17}
\ee
The first term in Eq. (\ref{c16}) is the standard expression of the Chandrasekhar mass limit whereas the second
term represents the modification from the quantum plasma Bohm potential. In order to investigate its effect in 
enhancing or reducing the standard mass limit one has to compute the first zero $x_1$ and the shape of the 
solution $y_3$ of the QMLE equation. 
\section{Approximate solution of the QMLE equation for general polytropic 
indices $\a$ for small values of $\eta_{\a}$}
Substituting $t=1/x$ the QMLE equation (\ref{c12}) reads 

\be
t^4{d^2\over dt^2}\l[y_{\a }-{\eta _{\a }t^4\over y^{\a /2}_{\a }}{d^2y^{\a /2}_{\a }\over dt^2}\r]
+y_{\a }^{\a }=0
\label{f1}
\ee
Dropping the index $\a$, i.e. $y(t)=y_{\a}(t)$ and $\eta _{\a} =\eta $, and using $u(t)=[y(t)]^{\a /2}$ 
the QMLE equation (\ref{f1}) reads

\be
{d^2\over dt^2}\bigl[y-{\eta t^4\over u}{d^2u\over dt^2}\bigr]+{u^2\over t^4}=0
\label{w1}
\ee
\subsection{First-order expansion in $\eta $}
For very small values of the quantum parameter $\eta \ll 1$ set 

\be
u(t)=u_c(t)+\eta \delta u
\label{w2}
\ee
where $u_c$ fulfills the standard Lane-Emden equation 

\be
{d^2u_c^{2/\a }\over dt^2}+{u_c^2\over t^4}=0
\label{w3}
\ee
The ansatz (\ref{w2}) implies

\bdm
y(t)=y_c(t)+\eta \delta y=\l(u_c+\eta \delta u\r)^{2/\a}
\edm
\be
\simeq u_c^{2/\a}+{2\eta \over \alpha }u_c^{(2-\a)/\a}\delta u
\label{w4}
\ee
so that to first order in $\eta $ 

\be
\delta u={\a \over 2}u_c^{(\a-2)/\a}\delta y
\label{w5}
\ee
Using Eqs. (\ref{w2}) and (\ref{w4}) we find for the QMLE equation (\ref{w1}) to first order in $\eta $ 

\be
{d^2\over dt^2}\bigl[y_c+\eta \delta y-{\eta t^4\over u_c}{d^2u_c\over dt^2}\bigr]+
{u_c^2+2\eta u_c\delta u\over t^4}=0
\label{w6}
\ee
Using Eqs. (\ref{w3}) and (\ref{w5}) we derive

\bdm
{d^2\over dt^2}\bigl[\delta y-{t^4\over u_c}{d^2u_c\over dt^2}\bigr]+{2\over t^4}u_c\delta u=
\edm
\be
{d^2\over dt^2}\bigl[\delta y-{t^4\over u_c}{d^2u_c\over dt^2}\bigr]+{\a \over t^4}u_c^{2(\a -1)/\a}\delta y=0
\label{w7}
\ee
which, for a known standard solution $u_c$, is a linear second-order differential equation for the deviation $\delta y$.

Both $u(t)$ and the solution $u_c(t)$ of the standard Lane-Emden equation fulfil the boundary conditions

\bdm
u_c(t=\infty )=1,\,\; {du_c\over dt}|_{t=\infty }=0,\,\: 
\edm
\be
u(t=\infty )=1,\,\; {du\over dt}|_{t=\infty }=0
\label{w8}
\ee
implying 

\be
\delta y(t=\infty )=0,\,\; {d\delta y\over dt}|_{t=\infty }=0
\label{w9}
\ee
With these boundary conditions, integrating equation (\ref{w7}) once over $t$, we obtain 

\be
{d\over dt}\l[\delta y-{t^4\over u_c}{d^2u_c\over dt^2}\r]
=2\int_t^\infty dq\; q^{-4}\delta u(q)u_c(q)
\label{w10}
\ee
\subsection{First-order quantum corrections to the Chandrasekhar mass and radius}
In terms of the variable $t=1/x$ the Chandrasekhar mass (\ref{c15}) reads 

\be
M(x_1)=4\pi \rho _c A^3{d\over dt}\l[y-{\eta t^4\over u}{d^2u\over dt^2}\r]_{t_1}
\label{w11}
\ee
where $t_1=1/x_1$ denotes the largest zero of the function $u(t_1)=0$. 
With the ans\"atze (\ref{w2}) and (\ref{w4}) we find to first order in $\eta $  

\bdm
M(x_1)=4\pi \rho _c A^3\l({dy_c\over dt}+\eta 
{d\over dt}\l[\delta y-{t^4\over u_c}{d^2u_c\over dt^2}\r]\r)_{t_1}
\edm
\be
=4\pi \rho _c A^3\l(({dy_c\over dt})_{t_1}+
2\eta \int_{t_1}^\infty dq\; {u_c(q)\delta u(q)\over q^4}\r)
\label{w12}
\ee
where we inserted equation (\ref{w10}). Eq. (\ref{w12}) represents the first-order correction in $\eta $ 
of the Chandrasekhar mass.

If $t_c$ denotes the largest zero of $u_c(t_c)=0$ we set $t_1=t_c+\delta t$. The boundary condition  

\bdm
0=u(t_1)=u_c(t_1)+\eta \delta u(t_1)
\edm
together with the Taylor expansion 

\bdm 
u_c(t_1)\simeq u_c(t_c)+({du_c\over dt})_{t_c}\delta t=({du_c\over dt})_{t_c}\delta t
\edm
then yields 

\be
\delta t=-\eta {\delta u(t_1)\over ({du_c\over dt})_{t_c}}\simeq 
-\eta {\delta u(t_c)\over ({du_c\over dt})_{t_c}}
\label{w13}
\ee
so that 

\be
t_1=t_c-\eta \tau,\;\;\; \tau= {\delta u(t_c)\over ({du_c\over dt})_{t_c}}
\label{w14}
\ee
Consequently, the dimensionless radius of the gas sphere 

\be
x_1={1\over t_1}\simeq {1\over t_c}+{\eta \tau \over t_c^2}
\label{w15}
\ee
is enhanced (diminished) in comparison to the Chandrasekhar radius $t_c^{-1}$ when $\tau >0$ ($\tau <0$).

With the Taylor expansion 

\bdm
y_c(t)\simeq y_c(t_c)+({dy_c\over dt})_{t_c}(t-t_c)+{1\over 2}({d^2y_c\over dt^2})_{t_c}(t-t_c)^2
\edm
we obtain 

\be
({dy_c\over dt})_{t_1}=({dy_c\over dt})_{t_c}-\eta \tau ({d^2y_c\over dt^2})_{t_c}
\label{w16},
\ee
so that the Chandrasekhar mass (\ref{w12}) becomes 

\be
M(x)=4\pi \rho _c A^3\l(({dy_c\over dt})_{t_c}+\eta B_c\r)
\label{w17}
\ee
where

\be
B_c=2\int_{t_c}^\infty dq\; {u_c(q)\delta u(q)\over q^4}-
{\delta u(t_c)\over ({du_c\over dt})_{t_c}}({d^2y_c\over dt^2})_{t_c}
\label{w18}
\ee
The first term in Eq. (\ref{w17}) represents the standard Chandrasekhar mass. The quantum correction scales 
linearly proportional to $\eta $. 

\subsection{Approximate solution of the differential equation (\ref{w7})}
It is well-known \cite{c58} that the standard Lane-Emden equation, i.e. Eq. (\ref{f1}) with $\eta _{\a }=0$, 
is analytically solvable only for the values $\a =0, 1, 5$ but not for the physically interesting case 
$\a =3$. However, for general values of $\a $ the standard Lane-Emden equation has 
the asymptotic solution \cite{c58}

\be
y_c(t)\simeq 1-{1\over 6t^2}+{\a \over 120t^4} 
\label{x1}
\ee
implying 

\be
u_c(t)=y^{\a /2}_c(t)\simeq 1-{\a \over 12t^2}+{\a ^2\over 240t^4}
\label{x2},
\ee
\be
u_c^{2-(2/\a)}\simeq 1-{\a -1\over 6t^2}+{\a (\a -1)\over 120t^4}
\label{x3},
\ee
\be
u_c^{1-(2/\a)}\simeq 1-{\a -2\over 12t^2}+{\a (\a -2)\over 240t^4}
\label{x3a},
\ee
and

\be
{d^2u_c\over dt^2}\simeq -{\a \over 2t^4}\l[1-{\a \over 6t^2}\r]
\label{x4},
\ee
respectively. Substituting 

\be
\delta y(t)=g(t)+{t^4\over u_c}{d^2u_c\over dt^2}\simeq 
g(t)-{\a \over 2}\l[1-{\a \over 12t^2}\r]
\label{x5}
\ee
yields for the differential equation (\ref{w7})

\be
{d^2g\over dt^2}+{\a \over t^4}u_c^{2(\a -1)/\a}g=-\a u_c^{1-(2/\a)}{d^2u_c\over dt^2}
\label{x6}
\ee
The boundary conditions (\ref{w9}) imply $g(t=\infty)=\a /2$. Inserting the asymptotics 
(\ref{x2}) - (\ref{x4}) yield for the differential equation (\ref{x6}) 

\bdm 
{d^2g\over dt^2}+{\a \over t^4}\l[1-{\a -1\over 6t^2}+{\a (\a -1)\over 120t^4}\r]g\simeq 
\edm
\be
{\a ^2\over 2t^4}\l[1-{3\a -2\over 12t^2}\r]
\label{x7}
\ee
which, at large $t$, has the asymptotic solution 

\be
g(t)\simeq {\a \over 2}-{\a ^3\over 480t^4}
\label{x8}
\ee
implying from equation (\ref{x5}) 

\be
\delta y(t)\simeq {\a ^2\over 24t^2}\l[1-{\a \over 20t^2}\r]
\label{x9}
\ee
From equation (\ref{w5}) we obtain accordingly  

\be
\delta u={\a \over 2}u_c^{(\a-2)/\a}\delta y
\simeq {\a ^3\over 48t^2}\l[1-{4\a -5\over 30t^2}\r]
\label{x10}
\ee
\subsection{Chandrasekhar mass and radius reductions}
Collecting terms we obtain for the quantum plasma correction (\ref{w18}) to the standard Chandrasekhar mass 
the quantity 

\be
B_c\simeq {\a ^2\over 8t_c^3}\l[1+{5-4\a \over 30t^2_c}\r]
\label{x11}
\ee
Likewise, the corrected normalised radius (\ref{w15}) is

\be
x_1\simeq {1\over t_c}\l[1+{\a ^2\eta \over 8}\l(1+{5-\a \over 30t_c^2}\r)\r]
\label{x12}
\ee
For $\a =3$ the corrected Chandrasekhar mass (\ref{w17}) thus becomes 

\be
M(x)=4\pi \rho _c A^3\l(({dy_c\over dt})_{t_c}+{9\eta \over 8t_c^3}
\l[1-{7\over 30t^2_c}\r]\r)
\label{x13}
\ee
whereas the corrected normalised radius (\ref{x12}) is
 
\be
x_1\simeq {1\over t_c}\l[1+{9\eta \over 8}\l(1+{1\over 15t_c^2}\r)\r]
\label{x14}
\ee
Strictly taken, expressions (\ref{x11}) - (\ref{x14}) are only valid for large values of $t_c\ll 1$ because 
the corrections have been derived from the corresponding asymptotic solutions. Nevertheless, we use them here 
for \cite{c58} $t_c=1/6.89685=0.145$, which is not large compared to unity, in order to obtain a first crude
estimate of the strength of the quantum corrections. We find that the quantum corrections reduce 
the standard Chandrasekhar mass by the negligibly small factor 

\be
B_c\eta =-3.73\cdot 10^3\eta =-3.4\cdot 10^{-35}(a_9\chi )^{-1}
\label{x15}
\ee
and enhance the normalised stellar radius by the negligibly small factor

\be
{9\eta \over 8}\l(1+{1\over 15t_c^2}\r)=4.69\eta =4.4\cdot 10^{-38}(a_9\chi )^{-1}
\label{x16}
\ee
\section {Summary and conclusions}
The proper quantum plasma treatment of the electron gas in degenerate stars such as white dwarfs provides an
additional quantum contribution to the electron pressure. We have calculated how this additional pressure term 
modifies the equation for hydrostatic equilibrium, resulting in the quantum modified Lane-Emden equation for
polytropic equation of states. The additional pressure term also modifies the expression for the limiting
Chandrasekhar mass of white dwarfs. We develop an approximate solution of the 
quantum modified Lane-Emden equation for general polytropic indices.We demonstrate that the 
quantum corrections reduce the standard Chandrasekhar mass by a negligibly small value of order 
${\cal O}(10^{-35})$ and enhance the stellar white dwarf radius by a negligibly small value 
of order ${\cal O}(10^{-38})$. Deviations from charge neutrality in the white dwarf electrron-ion plasma are of order 
${\cal O}(10^{-37})$.

Our study is preliminary in two important aspects. First we have used asymptotic expansions 
of our approximate solution for the QMLE equation for large values of the dimensionless variable $t$ 
to estimate the quantum corrections to the white dwarf mass and radius. Such can be improved in future work 
by a numerical solution of the QMLE equation for all values of $t$. Secondly, we have used a nonrelativistic
calculation for the additional quantum pressure term to combine it with the relativistic degeneracy electron
pressure term when estimating the corrections to the standard Chandrasekhar mass. This shortcoming is 
currently difficult to fix, as a fully relativistic kinetic theory of quantum plasmas is not completely 
fleshed out as of yet. 
\begin{acknowledgements} 
We are grateful to F. Haas, W. Moslem and P. K. Shukla for sharing their insights into 
quantum plasma physics.
\end{acknowledgements}

\begin{appendix}
\section{Quantum hydrodynamical equations}
According to Manfredi and Haas \cite{mh01} the quantum hydrodynamic equations of an electron-ion plasma 
in a 1-dimensional cartesian geometry are given by the Poisson equation  

\be
{d^2\Phi \over dz^2}=4\pi e\l[n-\, n_i\r]
\label{rb1},
\ee
the equation of continuity 

\be
{\partial n\over \partial t}+{\partial(nu)\over \partial z}=0
\label{rb2},
\ee
and the dynamical electron equation

\be
0={\partial u\over \partial t} +u{\partial u\over \partial z}+{1\over mn}{d(P_c+P_Q)\over dz}
+{e\over m}{d\Phi \over dz}
\label{rb3}
\ee
with the Fermi pressure $P_c$ and the Bohm pressure 

\be
P_Q={\hbar ^2\over 2m}\l[\l({\partial n^{1/2}\over \partial z}\r)^2-
n^{1/2}{\partial^2 n^{1/2}\over \partial z^2}\r]
\label{rb4}
\ee
$\Phi $ denotes the electrostatic potential and $n$ and $n_i$ refer to the electron and ion number densities, respectively. 
With the identity 

\bdm
{1\over nm}{dP_Q\over dz}={\hbar ^2\over 2m^2}{1\over n}{d\over dz}
\l[\l({\partial n^{1/2}\over \partial z}\r)^2-
n^{1/2}{\partial^2n^{1/2}\over \partial z^2}\r]
\edm
\be
=-{\hbar ^2\over 2m^2}{d\over dz}\l[{{d^2n^{1/2}\over dz^2}\over n^{1/2}}\r]
\label{rb5}
\ee
the dynamical electron equation in full 3-dimensional form reads 

\bdm
0={\partial \vec{u}\over \partial t} +(\vec{u}\cdot {\partial \over \partial \vec{x}})\vec{u}+
{e\over m}\nabla \Phi +{1\over mn}\nabla P_c-
\edm
\be
{\hbar ^2\over 2m^2}\nabla \l({\nabla ^2n^{1/2}\over n^{1/2}}\r)
\label{rb6}
\ee
\subsection{Quantum modified hydrostatic equilibria}
For static ($\vec{u}=0$) systems, after adding gravity, the electron dynamical equation reads 

\be
0=mn\vec{g}(\vec{x}) +en\nabla \Phi +\nabla P_c-
{\hbar ^2n\over 2m}\nabla \l({\nabla ^2n^{1/2}\over n^{1/2}}\r)
\label{rb7}
\ee
Because the ion contribution to the pressure is negligibly small, the corresponding ion equation reads 

\be
0\simeq m_in_i\vec{g}(\vec{x}) -Zen_i\nabla \Phi 
\label{rb8}
\ee
Adding equations (\ref{rb7}) and (\ref{rb8}) and using the charge neutrality condition $Zn_i=n$ 
eliminates the electrostatic potential leaving 

\be
0=(mn+m_in_i)\vec{g}(\vec{x})+\nabla P_c-
{\hbar ^2n\over 2m}\nabla \l({\nabla ^2n^{1/2}\over n^{1/2}}\r)
\label{rb9}
\ee
With the mass density $\rho =m_in_i+mn\simeq m_in/Z$ we derive 

\be
0=\vec{g}(\vec{x})+{1\over \rho }\nabla P_c-
{\hbar ^2Z\over 2mm_i}\nabla \l({\nabla ^2\rho ^{1/2}\over \rho ^{1/2}}\r)
\label{rb10}
\ee
which is identical to Eq. (\ref{c1}). Under charge neutrality the electric field $\vec{E}=-\nabla \Phi $ 
does not enter the equation for stellar hydrostatic equilibrium.
\subsection{Electric field equation and charge neutrality state}
Multiplying equation (\ref{rb8}) with $Z$ and subtracting equation (\ref{rb7}) 
yields for the electrostatic potential \cite{km75} 

\bdm
\l(Z^2en_i+en\r)\nabla \Phi =\l(Zm_in_i-mn\r)\vec{g}(\vec{x})
\edm
\be
-\nabla P_c
+{\hbar ^2n\over 2m}\nabla \l({\nabla ^2n^{1/2}\over n^{1/2}}\r)
\label{rb11}
\ee
Because of the stronger action of the gravitational field on the plasma ions as compared to the plasma electrons, a 
charge separtion results that induces a nonzero electrostatic potential and electric field. Assuming only small
perturbations of the charge neutrality condition $Zn_i\simeq n$ then yields 

\bdm
\vec{E}=-\nabla \Phi \simeq -{m_i\over e(Z+1)}\vec{g}(\vec{x})
+{\nabla P_c\over e(Z+1)n}
\edm
\be
-{\hbar ^2\over 2me(Z+1)}\nabla \l({\nabla ^2\rho ^{1/2}\over \rho ^{1/2}}\r)
=-{m_i\over eZ}\vec{g}(\vec{x})
\label{rb12}
\ee
where we have inserted Eq. (\ref{rb9}). This electric field provides a net space charge density $\sigma $ that 
can be estimated by calculating the divergence of Eq. (\ref{rb12}) as

\be
\nabla \cdot \vec{E}=4\pi \sigma =4\pi e(Zn_i-n)=-{m_i\over eZ}\nabla \cdot \vec{g}
\label{rb13}
\ee
With $\nabla \cdot \vec{g}=-4\pi G\rho $ and $\rho =m_in_i$ we obtain the positive space charge density

\be
\sigma =e(Zn_i-n)={Gm_i^2n_i\over eZ}={Gm_i^2n\over eZ^2}
\label{rb14}, 
\ee
that has to be screened by an appropriately negative space charge density on the surface of the star \cite{schl50}.
As relative deviation from the overall charge neutrality state 
we obtain 

\be
{Zn_i-n\over Zn_i+n}={Zn_i-n\over 2Zn_i}={Gm_i^2\over 2e^2Z^2}=1.01\cdot 10^{-37}
\label{rb15}
\ee
if we use $Z=2$ and $m_i=m_p$. The deviation from the overall charge neutrality state is negligibly small.

\section{Relativistic generalisation of the quantum modified Lane-Emden equation}
Assuming only isotropic stresses, the mixed components of the energy-momentum tensor of a 
spherically symmetric distribution of fluid obeying a polytropic equation of state in a co-moving system read 

\begin{equation}
T_1^{\;1} = T_2^ {\;2} = T_3^ {\;3} = -P \; \; \;\; \; \;\; \; \;\; \; \;\; \; \; T_0^{\;0} = \epsilon = \rho c^2 \, ,
\end{equation}\\
where $P$ is the pressure and $\epsilon = \rho c^2$ is the energy density. Using a spherical form of the metric 
at rest with respect to the fluid distribution

\begin{equation}
ds^2 = e^{\nu(r)} c^2 dt^2 - r^2 \bigl(d\theta^2 + \sin^2{\theta} d\phi^2 \bigr) - e^{\lambda(r)} dr^2
\end{equation}\\
we obtain the time-independent gravitational equations \cite{ll62}

\begin{equation}\label{REL0}
e^{-\lambda} \biggl(\frac{1}{r} \frac{d\nu}{dr} + \frac{1}{r^2} \biggr) - \frac{1}{r^2} = \frac{8 \pi G}{c^4} P
\end{equation}

\begin{equation}\label{REL00}
e^{-\lambda} \biggl(\frac{1}{r} \frac{d\lambda}{dr} - \frac{1}{r^2} \biggr) + \frac{1}{r^2} = \frac{8 \pi G}{c^4} \rho c^2
\end{equation}

\begin{equation}
e^{-\lambda} \biggl(\frac{d^2\nu}{dr^2} + \frac{1}{2} \biggl[ \frac{d\nu}{dr}\biggr]^2 + \frac{1}{r} \biggl[ \frac{d\nu}{dr} - \frac{d\lambda}{dr}\biggr] - \frac{1}{2} \frac{d\nu}{dr} \frac{d\lambda}{dr}  \biggr) = \frac{16 \pi G}{c^4} P
\end{equation}\\
Combining these equations we derive the expression

\begin{equation} \label{REL1}
\frac{dP}{dr} + \frac{1}{2} \bigl(P + \rho c^2 \bigr) \frac{d\nu}{dr} = 0
\end{equation}
which is the general relativistic analogue of the classical form of equation (\ref{c2}). 
Introducing the quantum correction $P_Q$ in the expression for the pressure $P = P_c + P_Q$ (see equation (A18) and (A20)) equation (B6) yields\\

\begin{equation}\label{REL2}
\begin{split}
&\frac{1}{\rho(r)}\frac{dP_c}{dr} - \frac{\hbar^2}{2 mm_i} \frac{d}{dr} \biggl( \frac{\nabla^2_r \rho^{1/2}(r)}{\rho^{1/2}(r)}\biggr) + \frac{1}{2 \rho(r)} \biggl(P_c + \\ \\
&\frac{\hbar^2}{2 mm_i} \biggl[ \biggl( \frac{d\rho^{1/2}(r)}{dr}\biggr)^2 - \rho^{1/2}(r) \nabla^{2}_r \rho^{1/2}(r)\biggr]+ \rho c^2 \biggr) \frac{d\nu}{dr} = 0
\end{split}
\end{equation}\\
Using the polytropic equation of state (\ref{c8}) and the substitutions 
$\displaystyle{w(\zeta) = \frac{1 - e^{-\lambda}}{2 (\alpha + 1) \sigma} \,\zeta}$, where $A$ is defined 
by expression (\ref{c11}), $\zeta = r/A$ and the constants 
$\displaystyle{\chi_{\alpha} = \frac{\hbar^2}{2 mm_i A^2 c^2 (1 + \sigma y_{\alpha})}}$, 
$\sigma = K \rho^{1/\alpha}_c / c^2$ and inserting equation (\ref{REL2}) into (\ref{REL0}) we derive 
neglecting terms of order $\mathcal{O}(\hbar^4/A^4)$\\

\begin{equation}\label{REL3}
\begin{split}
& w +\sigma y_{\alpha} \zeta \frac{dw}{d\zeta} + \frac{1 - 2 (\alpha +1) \sigma w / \zeta}{1 + \sigma y_{\alpha}} \, \, \zeta^2 \frac{dy_\alpha}{d\zeta} = \\ \\
&  \chi_{\alpha} \Biggl(y^{-\alpha}_ \alpha \biggl(  (2 \sigma y_\alpha + 1) \zeta \frac{dw}{d\zeta} + w \biggr) \biggl[ y^{\alpha/2}_ \alpha \nabla^2_\zeta y^{\alpha/2}_ \alpha  - \biggl( \frac{dy^{\alpha/2}_\alpha}{d\zeta}\biggr)^2\biggr] \\ \\
&+ \frac{1 - 2 (\alpha +1) \sigma w / \zeta}{\sigma (\alpha + 1)} \, \, \zeta^2 \frac{d}{d\zeta}\biggl( \frac{\nabla^2_\zeta y^{\alpha/2}_\alpha}{y^{\alpha/2}_\alpha} \biggr) \Biggr)
\end{split}
\end{equation}\\
Equation (\ref{REL00}) yields with the definitions and substitutions given previously\\
\begin{equation}\label{REL4}
\frac{dw}{d\zeta} = \zeta^2 y^\alpha_\alpha
\end{equation}\\
In the non-relativistic limit the general relativistic quantum system (\ref{REL3}) and (\ref{REL4}) reduces 
to equation (\ref{c12}), while for $\hbar \rightarrow 0$ it reduces to the general 
relativistic Lane-Emden equation of stellar structure \cite{t64}.

\end{appendix}

\end{document}